\documentclass[twocolumn,prd,showpacs,preprintnumbers,nofootinbib]{revtex4-1}
\usepackage{amsmath,amssymb,pgf,graphicx,color,url}
\newcommand{\be}{\begin{equation}}
\newcommand{\ee}{\end{equation}}

\definecolor{SIENA}{rgb}{0.772,0.541,0.243}

\usepackage[unicode=true,pdfusetitle,
bookmarks=true,bookmarksnumbered=true,bookmarksopen=true,bookmarksopenlevel=1,
breaklinks=false,pdfborder={0 0 0},backref=false,colorlinks=true]
{hyperref}
\hypersetup{
	citecolor=blue,filecolor=blue,linkcolor=blue,urlcolor=blue}

\begin{document}

\preprint{CERN-TH-2016-034}
\preprint{INR-TH-2016-005}

\title{On constraining the speed of gravitational waves following GW150914}

\author{
Diego Blas$^{a}$, 
Mikhail \,M.\,Ivanov$^{b,c,d}$, Ignacy Sawicki$^{e}$
and Sergey \,Sibiryakov$^{a,b,c}$}
\affiliation{
$^a$ Theoretical Physics Department, CERN, CH-1211 Geneva 23, Switzerland\\
$^b$ FSB/IPHYS/LPPC, \'Ecole Polytechnique F\'ed\'erale de Lausanne, \normalsize\it CH-1015, Lausanne, Switzerland \\
$^c$ Institute  for Nuclear Research of the Russian Academy of Sciences, \normalsize\it 117312 Moscow, Russia\\
$^d$ Faculty of Physics, Moscow State University, \normalsize\it 119991 Moscow, Russia \\
$^e$ D\'epartment de Physique Th\'eorique and Center for Astroparticle Physics,
Universit\'e de Gen\`eve, Quai E. Ansermet 24, CH-1211 Gen\`eve 4, Switzerland
}

\begin{abstract}
We point out that the observed time delay between the detection of the
signal at the Hanford and Livingston LIGO sites from the gravitational
wave event GW150914 places an upper bound on the speed of propagation
of gravitational waves, $c_{gw}\lesssim 1.7$ in the units of speed of
light.  Combined with the lower bound from the absence of
gravitational Cherenkov losses by cosmic rays that rules out most of
subluminal velocities, this gives a model-independent double-sided
constraint $1\lesssim c_{gw}\lesssim 1.7$. We compare this result to
model-specific constraints from 
pulsar timing and cosmology.
\end{abstract}

\maketitle

\section{Introduction}

The recent discovery of gravitational waves (GWs) by  the LIGO collaboration \cite{Abbott:2016blz} opens a window for testing fundamental properties of gravitation \cite{TheLIGOScientific:2016src,Calabrese:2016bnu,Berti:2015itd}. In this short note, we point out that these results can be used to bound the speed of propagation of GWs in a model-independent manner, giving complementary, if weaker, constraints to model-specific ones already in the literature.

If one considers that GWs propagate as free waves, one can
parameterize the dispersion relation of their frequency $\omega$ and
momentum $k=|\bf{k}|$ by the generic formula (we restrict to theories
where gravitational waves are described by equations with 
two time-derivatives and assume rotational invariance),
\be
\label{eq:dispersion}
\omega^2=m_{gw}^2+c^2_{gw}k^2 + \alpha_{(4)}\frac{k^4}{\Lambda^2}+.... \,,
\ee
where we have introduced a mass $m_{gw}$ term, the possibility of a
speed of propagation $c_{gw}$ different from\footnote{We work in the
  units where the speed of light is equal to 1.} $1$ for modes $k\ll
\Lambda$ and a high scale $\Lambda$ beyond which dispersive effects
become relevant. The modification of the velocity and the dispersion 
appear in some approaches to  
quantum gravity \cite{Liberati:2013xla,Blas:2014aca}. 
The mass $m_{gw}$ has already been constrained by the LIGO
collaboration in \cite{TheLIGOScientific:2016src}: $m_{gw}\leq 1.2\times 10^{-22} \
\mathrm{eV}$ by studying the arrival time of different frequency
components of the signal that has traveled across $\sim 400 \ \mathrm{Mpc}$.

We will assume that the high-energy scale $\Lambda$ is much higher
than the characteristic frequency of the signal $\sim 100 \
\mathrm{Hz}$, and that $m_{gw}$ satisfies the LIGO bound. LIGO
Collaboration has not placed a constraint on $c_{gw}$: in
\cite{TheLIGOScientific:2016src}, $c_{gw}=1$ is assumed in order to
allow for the localization of event GW150914 in the sky. In this note,
we point out that independently of the degeneracy with the source
direction, 
GW150914 can be used to place the first {\it model independent upper} bound on $c_{gw}$.

\section{Existing bounds on $c_{gw}$}

We first remind that the interesting region to constrain corresponds to 
\be
\label{eq:Ch}
1-c_{gw} \lesssim 10^{-15}.
\ee
This is a conservative bound arising from the absence of gravitational
Cherenkov radiation allowing for the unimpeded propagation of
high-energy cosmic rays across our galaxy \cite{Moore:2001bv} (see
\cite{Caves:1980jn} for an earlier study and
also \cite{Kiyota:2015dla} for the study with generic dispersion relations). 
Note that another lower bound $1-c_{gw} \lesssim 10^{-2}$ can be obtained from pulsar timing
\cite{Baskaran:2008za}. 
Though weaker, this bound directly constrains the speed of classical gravitational waves and is independent of the microscopic interactions of gravitons with high-energy particles.

Second, regarding upper bounds, let us note
that for theories where Lorentz invariance is broken at the fundamental
level, 
there is no theoretical argument (or pathology) against signals propagating faster than light \cite{Babichev:2007dw,Blas:2014aca}. Furthermore, these theories 
provide a set-up where gravity can be potentially 
quantized using the standard framework of
quantum field theory in 4-dimensions \cite{Horava:2009uw,Barvinsky:2015kil}. Thus, a measurement of $c_{gw}$ is a concrete application of LIGO results to test the  ideas of quantum gravity.
 
Upper bounds on $c_{gw}$ 
can and have been obtained from various astrophysical and cosmological tests. 
However, these bounds are model dependent. For instance, in the case
of Lorentz-violating theories, the bounds from radiation damping in
binary systems imply 
$c_{gw}-1\lesssim 10^{-2}$ \cite{Yagi:2013qpa}. Similar constraints have
been derived is scalar-tensor theories \cite{Jimenez:2015bwa} and
using cosmology\footnote{
In principle, one can use the fact that $c_{gw}\neq 1$ is related to
the presence of gravitational slip to eventually produce stronger
bounds from cosmology \cite{Saltas:2014dha, Amendola:2014wma}.} 
\cite{Audren:2014hza,Bellini:2015xja}. Forecasts for different
constraints using advanced detectors can be found in
\cite{Mirshekari:2011yq}.
See also 
\cite{Raveri:2014eea}
for the constraints on the speed of cosmological gravitational waves
with future CMB instruments.

\section{Upper bounds from GW150914}

There is clearly an interest in setting an upper bound on $c_{gw}$. Let us recall that in \cite{Abbott:2016blz,TheLIGOScientific:2016src} this was not done since it was assumed that $c_{gw}=1$ and the difference in the time of arrival of the signal to the different interferometers of LIGO was used to localize the event. 

\begin{figure}
\begin{center}
\includegraphics[width=.45\textwidth]{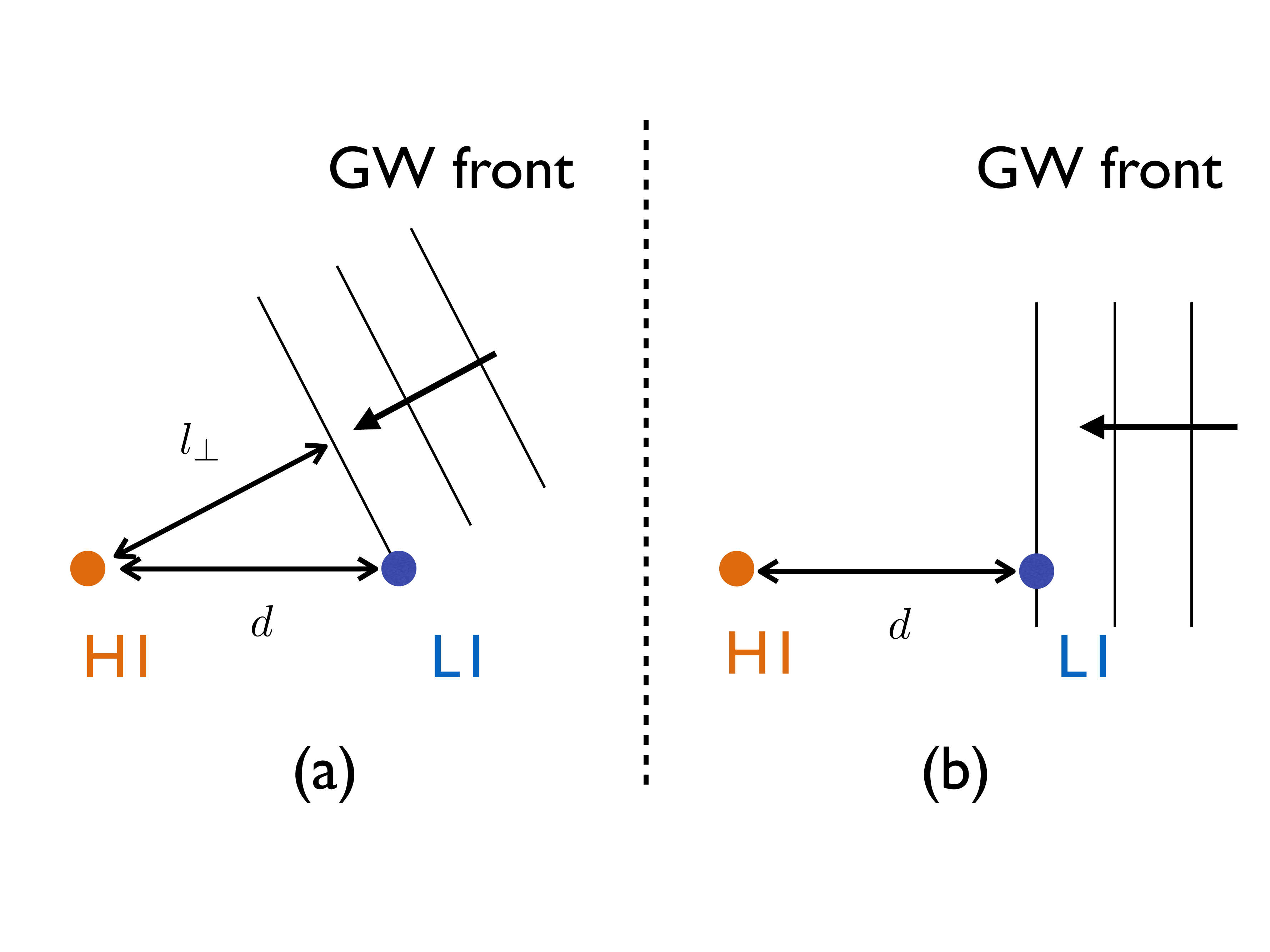}
\caption{\label{fig:LIHI}
$(a)$ incidence of GW from a generic direction. 
$(b)$ orientation giving the maximal time delay.}
\end{center}
\end{figure}

One can also take a different view. We use the fact that the two LIGO
sites at Livingston (L1) and Hanford (H1) separated by the distance of 
$d=10$~ms light travel time 
have detected the signal with the
time shift of $\Delta t=6.9^{+0.5}_{-0.4}$~ms \cite{Abbott:2016blz}. 
This time delay is equal to  
the projection $l_\perp$ of the intersite distance $d$ on the
direction perpendicular to the gravitational
wavefront (see Fig.~\ref{fig:LIHI}), divided by $c_{gw}$,
\be
\Delta t=l_\perp/c_{gw}\;.
\ee
Independently of the arrival
direction of the GW, $l_\perp$ cannot be larger than the
intersite distance $d$ itself  
which gives the bound
\be
c_{gw}\Delta t \leq d\;.
\ee
Substituting conservatively the minimal value of $\Delta t$ within
two-sigma deviation from the mean, we get,
\be
c_{gw}< 1.7 \label{eq:bound}\;.
\ee

\section{Discussion}

We have shown how our very naive reinterpretation of the analysis
of the detection of GW150914 sets the first direct  
bound on the speed of propagation of GWs, eq.~\eqref{eq:bound}. This
bound 
complements the lower bound coming from observations of high-energy cosmic rays \eqref{eq:Ch}.

Our constraint is already interesting and yet 
very conservative. We believe it can be improved 
by considering
other features of the event, such as orientations of the
detectors and the resulting antenna patterns, 
the amplitudes of the waveforms measured at the two sites 
or more information about the position of the source in the sky.

We have made the assumption that the change in the emission process as
a result of a modification of gravity would not affect the measurement
of the time delay between the two waveforms significantly, even if it
would affect the determination of the source parameters. One can
envisage two approaches to relax this assumption. The first is to develop
complete 
numerical simulations of compact binary coalescence 
in existing theories predicting deviations of
$c_{gw}$ from 1. Alternatively, one can focus just on the propagation
of the GW and implement a data analysis that would disentangle the 
measurement of the time delay from the model of the GW
emission. 

We believe that the results obtained by pursuing both these directions
will be of great value to fundamental physics.

\paragraph*{Acknowledgements} This work was supported 
by the Swiss National Science Foundation
(M.I. and S.S.), the RFBR grant 14-02-00894 (M.I.),  and by the Maria Sklodowska-Curie Intra-European Fellowship Project ``DRKFRCS'' (I.S.).

\end{document}